\documentclass[aps, twocolumn,
superscriptaddress,showpacs,showkeys,amsmath,amssymb,floatfix]{revtex4}
\usepackage{color}
\usepackage{graphicx}
\usepackage{epsfig}
\usepackage{dcolumn}
\usepackage{bm}
\usepackage{mathtools}
\usepackage{amssymb}
\usepackage{calrsfs}
\DeclareMathAlphabet{\pazocal}{OMS}{zplm}{m}{n} 
\usepackage{amsmath}
\usepackage{subfigure}
\newcommand{\be}{\begin{equation}}
\newcommand{\ee}{\end{equation}}
\newcommand{\bea}{\begin{eqnarray}}
\newcommand{\eea}{\end{eqnarray}}

\newcommand{\tK}{{\tilde K}}
\newcommand{\tQ}{{\tilde Q}}
\newcommand{\tS}{{\tilde S}}
\newcommand{\tF}{{\tilde F}}

\newcommand{\pro}{\partial}

\newcommand{\ba}{\begin{array}}
\newcommand{\ea}{\end{array}}

\newcommand{\nn}{\nonumber}

\newcommand{\Da}{\mathcal{D}}

\newcommand{\Ba}{\mathcal{B}}

\begin{document}
\title{On microscopic structure of the QCD vacuum}
\bigskip
\author{D.G. Pak}
\affiliation{Institute of Modern Physics, Chinese Academy of Sciences, 
Lanzhou 730000, China} 
\affiliation{Sogang University, Seoul 121-742, Korea}
\affiliation{Chern Institute of Mathematics, Nankai University,
Tianjin 300071, China}
\author{Bum-Hoon Lee}
\affiliation{Sogang University, Seoul 121-742, Korea}
\affiliation{Asia Pacific Center of Theoretical Physics,
Pohang, 790-330, Korea}
\author{Youngman Kim}
\affiliation{Rare Isotope Science Project, Institute for Basic Science,
Daejeon 305-811, Korea}
\author{Takuya Tsukioka}
\affiliation{School of Education, Bukkyo University, Kyoto 603-8301, Japan}
\author{P.M. Zhang}
\affiliation{Institute of Modern Physics, Chinese Academy of Sciences,
Lanzhou 730000, China}
\begin{abstract}
We propose a new class of regular stationary axially symmetric solutions 
in a pure QCD which correspond to monopole-antimonopole pairs at macroscopic
scale. The solutions represent vacuum field configurations which are locally stable 
against quantum gluon fluctuations in any small space-time vicinity. This implies that
the monopole-antimonopole pair can serve as a structural element in microscopic 
description of QCD vacuum formation through the monopole pair condensation.
\end{abstract}
\maketitle

\section{Introduction}

One of most attractive mechanisms of confinement is based on  idea
 that the vacuum in quantum chromodynamics (QCD) vacuum represents a dual superconductor
which is formed due to condensation of color magnetic
monopoles \cite{nambu74,mandelstam76,polyakov77,thooft81}.
Realization of this idea in the framework of a rigorous
theory has not been possible due to two principal difficulties: 
the first one, the color monopoles as physical particles have not been found 
in experiment, and so far there is no strict theoretical evidence
of their existence in the real QCD (except unphysical singular 
monopole solutions).
The second obstacle represents a long-standing problem  of vacuum stability 
since late 70s when it was shown that Savvidy-Nielsen-Olesen vacuum is unstable \cite{savv, N-O}. 
There have been numerous attempts to resolve this issue.
Unfortunately, most of vacuum models lack the local vacuum stability 
at microscopic space-time scale. Besides, the known approaches to vacuum problem
are based on assumption that vacuum field is static. Such an assumption is not consistent 
with the quantum mechanical principle
which implies that elementary vacuum fields (vortices, monopoles etc.) should 
vibrate at the microscopic level \cite{niel-oles1,amb-oles1}. 

In the present Letter we explore a novel approach for a 
microscopic description of the QCD vacuum based on stationary
vacuum field configurations. There are several no-go theorems which
impose strict limitations on the existence of finite energy static and stationary solutions in a pure
Yang-Mills theory \cite{derr64, deser76,pagels77,coleman77}.
To overcome the restrictions imposed by these no-go theorems,
it was proposed in past that classical stationary non-solitonic inifnite energy wave solutions 
with a finite energy density may correspond to quasi-particles or to vacuum
states in the quantum theory \cite{jackiw77}.
Recently an example of a regular stationary spherically symmetric solution
with a finite energy density in a pure QCD has been proposed \cite{ijmpa2017}.
The solution represents a system of interacting static Wu-Yang monopole
and time dependent off-dagonal gluon field. It has been proved that
such a solution is stable against quantum gluon fluctuations \cite{prd2017}. 
We generalize these results to the case of axially-symmetric fields 
 and construct a new class of regular axially symmetric periodic wave type solutions
with a finite energy density in a pure $SU(2)$ and $SU(3)$  QCD.
The solutions possess an intrinsic mass parameter which
determines a microscopic space-time scale. The averaging over the time period
implies that the proposed solutions can be treated as non-topological non-Abelian
monopole-antimonopole pairs. The most important result is that our solutions
are locally stable under quantum gluon fluctuations in any small vicinity of each space point.
The solutions can be served as structural elements of the QCD vacuum 
formed through the monopole pair condensation in analogy with the Cooper electron pair
condensation in the ordinary superconductor. 

\section{Axially-symmetric stationary solutions in $SU(2)$ QCD}

We start with a standard Lagrangian of a pure $SU(N)$ Yang-Mills theory 
and corresponding equations of motion
\bea
&&  {\cal L}_0 = -\dfrac{1}{4} F_{\mu\nu}^a F^{a\mu\nu}, \nn \\
&&  (D^\mu F_{\mu\nu})^a =0,\label{eqs0}
\eea
where $(a=1,2,.., N)$ denote the color indices, and $(\mu,\nu=r,\theta,\varphi,t)$ 
are the space-time indices in the spherical coordinates.
Let us first consider stationary solutions in a pure $SU(2)$ QCD.  
One can generalize a static 
axially symmetric Dashen-Hasslacher-Neveu (DHN) ansatz \cite{DHN} 
as follows
\bea
A_r^2&=& K_1(r,\theta,t),~~A_\theta^2= K_2(r,\theta,t), ~~A_\varphi^3= K_3(r,\theta,t), \nn \\
A_\varphi^1&=& K_4(r,\theta,t), ~~~A_t^2= K_5(r,\theta,t).  \label{genDHN}
\eea
The ansatz leads to a system of five differential equations
which are invariant under residual $U(1)$ gauge transformations 
with a gauge parameter  $\lambda(r,\theta,t)$ \cite{Manton78,RR,KKB} 
\bea
K'_1&=&K_1\pro_r \lambda,~~K'_2=K_2+\pro_\theta \lambda,~~K'_5=K_5+\pro_t \lambda, \nn \\
K_3'&=&K_3 \cos  \lambda+K_4\sin  \lambda, \nn \\
K_4'&=&K_4 \cos  \lambda-K_3 \sin  \lambda. \label{residual}
\eea
To fix the residual symmetry we choose a Lorenz gauge by introducing 
the gauge fixing terms
\bea
{\cal L}_{g.f.}^{\small {SU(2)}}&=&-\dfrac{1}{2} (\pro_r K_1+\dfrac{1}{r^2} \pro_\theta K_2 -\pro_t K_5)^2. \label{gfterm}
\eea

We apply a method used in solving equations for the
sphaleron solution \cite{RR,KKB}.
First, we decompose the functions $K_i$ in the Fourier series
\bea
K_{i=1-4}(r,\theta,t)&=&\delta_{i4} C_4+\nn \\
&&\sum_{n=1,2,...} \tilde K_{i}^{(n)}(nMr,\theta)  \cos (nMt), \nn \\
K_5(r,\theta,t)&=&\sum_{n=1,2,...} \tilde K_{5}^{(n)}(nMr,\theta)  \sin(nMt),  \label{seriesdec}
\eea
where $C_4$ is an arbitrary number, $M$ is a mass scale parameter which defines a class
of conformally equivalent field configurations due to the presence of scaling invariance in a pure QCD,
$(r\rightarrow Mr, t \rightarrow Mt)$. 
The structure of the series expansion (\ref{seriesdec}) implies
that all time averaged color electric components of
the field strength $F_{\mu\nu}^a$ vanish identically. 
Substituting the series decomposition  truncated at a finite order $n_f$ into the
classical action, and performing integration over the time period,
one results in a reduced action which implies Euler equations for the 
coefficient functions $\tilde K_{i}^{(n\leq n_f)}(r,\theta)$. 
The obtained reduced equations for the modes $\tilde K_{i}^{(n\leq n_f)}(r,\theta)$ 
represent a well-defined system of elliptic partial differential equations
which can be solved by applying standard numeric recipes.

\begin{figure}[h!]
\centering
\subfigure[~]{\includegraphics[width=42mm,height=30mm]{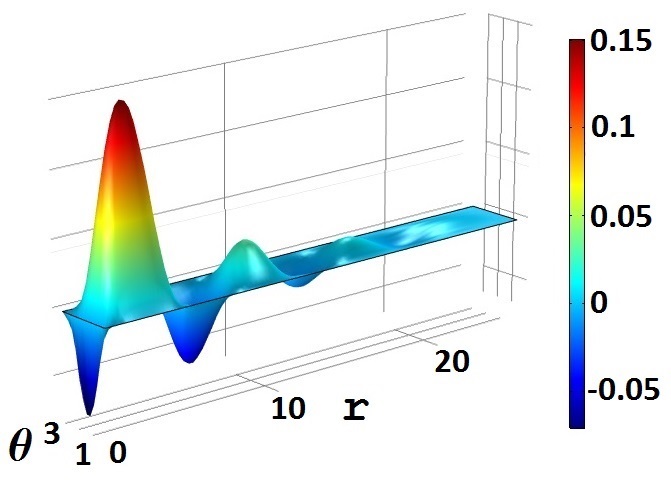}}
\subfigure[~]{\includegraphics[width=42mm,height=30mm]{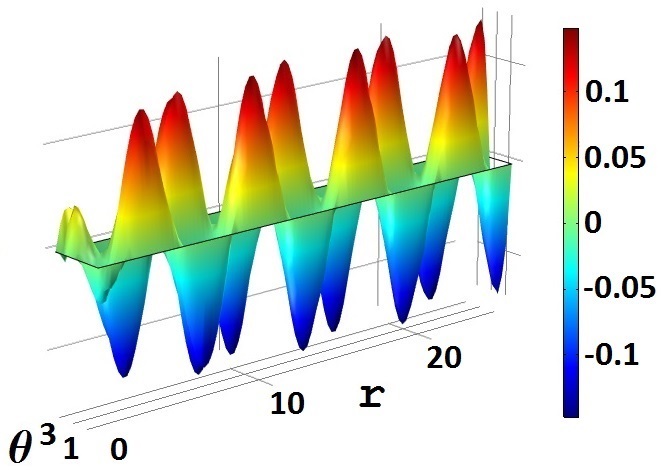}}
\subfigure[~]{\includegraphics[width=42mm,height=30mm]{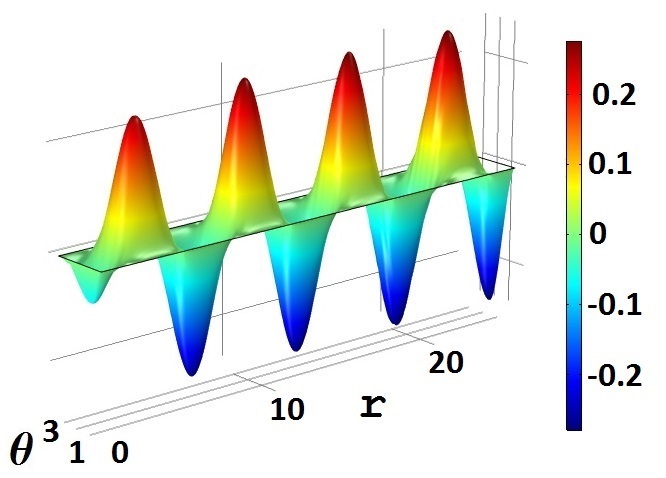}}
\subfigure[~]{\includegraphics[width=42mm,height=30mm]{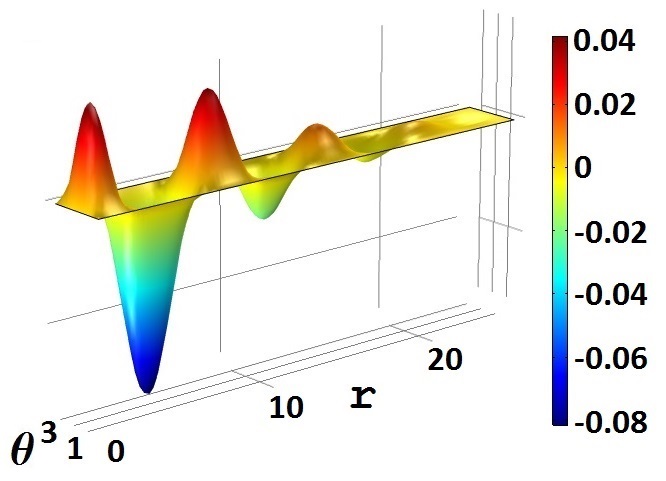}}
\subfigure[~]{\includegraphics[width=42mm,height=30mm]{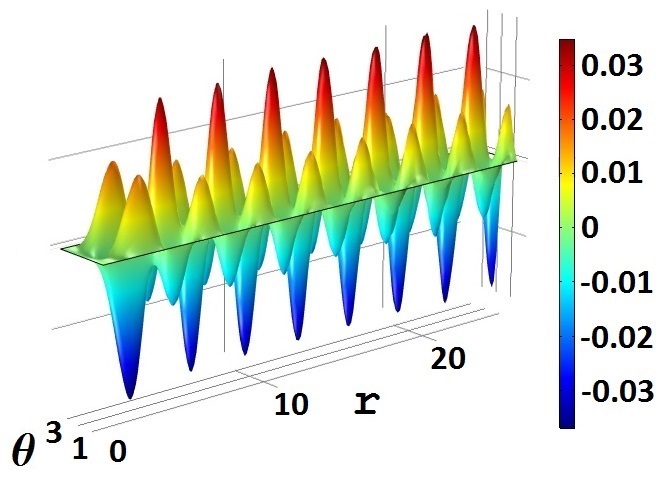}}
\hfill
\subfigure[~]{\includegraphics[width=42mm,height=32mm]{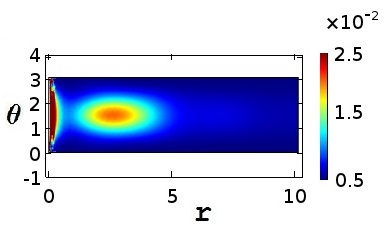}}
\caption[fig1]{Solution profile functions in the leading and sub-leading order: (a) $\tK_{1}^{(1)}$;
(b)  $\tK_{2}^{(1)}$; (c)  $\tK_{3}^{(1)}$; (d)   $\tK_{5}^{(1)}$; (e)  $\tK_{4}^{(2)}$; 
(f) the energy density plot ($C_4=2,g=1,M=1$).}\label{Fig1}
\end{figure}

To solve the equations for $\tilde K_{i}^{(n\leq n_f)}(r,\theta)$
we impose the following Dirichlet boundary conditions 
\bea
&&\tK_i^{(n)}(r,\theta)|_{r=0}=0, ~~~\tK_i^{(n)}(r,\theta)|_{\theta=0, \pi}=0.
\eea
Asymptotic behavior at ($r\rightarrow\infty$)  of the functions $\tK_i^{(n)}$ 
is determined by the equations of motion as follows
\bea
&&\tK_1^{(n)}\simeq a_1^{(n)}(\theta) \dfrac{ \sin(n Mr)}{r^2},~~~
                     \tK_5^{(n)}\simeq a_5^{(n)}(\theta)  \dfrac{ \cos(nM r)}{r^2}, \nn \\
&&\tK_{2,3,4}^{(n)}\simeq a_{2,3,4}^{(n)} (\theta) \cos(nM r), \label{asym1}
\eea
where $a_i^{(n)}(\theta)$ are periodic angle functions. 
We choose the lowest angle modes for 
$a_i^{(n)}(\theta)$ consistent with the finite energy density condition, 
in particular, in the leading order one has
$a_{i\neq 2}^{(1)}(\theta)=c_i \sin^2\theta$,
$a_{2}^{(1)}(\theta)=c_2\sin(2\theta)$. 
With this one can solve numerically the equations 
up to the sixth order of series decomposition
in the numeric domain $(0\leq r \leq L, 0\leq \theta  \leq \pi)$.
The obtained numeric solution implies that even order coefficient
functions $\tilde K_{1,2,3,5}^{(n=2k)}$ 
and odd order functions  $\tilde K_{4}^{(n=2k-1)}$ vanish identically. 
A typical solution in the leading and subleading order approximation
is presented in Fig. \ref{Fig1}.

One has fast convergence for the obtained numeric solution.
The solution profile functions of the third and fourth order, $\tilde K_{i}^{(n=3,4)}$,
provide corrections less than $2\%$ by a norm with respect to
the solution obtained in the leading and subleading order.
The fifth and sixth order corrections are less than $0.1\%$.
The energy density has an absolute maximum at the origin and 
a local maximum along the torus center line in the plane $\theta=\pi/2$.
For large values of the radius of the sphere enclosing the numeric domain, the
energy density decreases as $\dfrac{1}{R^2}$. 

In the leading order one has non-vanishing time averaged 
color magnetic fields
$\langle F_{\theta\varphi}^1 \rangle_t\simeq \tK_2^{(1)}\tK_3^{(1)}$ and
$\langle F_{r\varphi}^{1}\rangle_t\simeq \tK_1^{(1)}\tK_3^{(1)}$,
which create magnetic fluxes corresponding to 
a pair of non-topological monopole and antimonopole 
located at the origin. 
It is unexpected that a regular monopole-antimonopole pair solution
in the limit of zero distance between the monopoles does exist in the real QCD.
A brief overview of the structure 
of non-Abelian monopole-antimonopole fields and details of the numeric solution
are presented in \cite{SM}.


\section{Axially-symmetric stationary solutions in $SU(3)$ QCD}

We are looking for essentially $SU(3)$ field configurations
which do not reduce to embedded $SU(2)$ solutions. 
We start with a more general Lagrangian which includes additional gauge fixing terms
\bea
{\cal L}_{gen}&=& {\cal L}_0 - \sum_{a=2,5,7} \dfrac{\alpha_a}{2} (\pro_r A^a_1
+\dfrac{1}{r^2} \pro_\theta A^a_2 -\pro_t A_4^a)^2,\nn \\
&& \label{Lagrgen}
\eea
where $\alpha_a$ are arbitrary real numbers.
An axially symmetric ansatz contains the following
non-vanishing components of the gauge potential 
corresponding to three $I,U,V$-type $SU(2)$ subgroups of $SU(3)$
\bea
A_1^2&=&K_1,~~A_2^2=K_2,~~A_3^1=K_4,~~A_4^2=K_5, \nn \\
A_1^5&=&Q_1,~~A_2^5=Q_2,~~A_3^4=Q_4,~~A_4^5=Q_5, \nn \\
A_1^7&=&S_1,~~~A_2^7=S_2,~~A_3^6=S_4,~~~A_4^7=S_5, \nn \\
A_3^3&=&K_3,~~~A_3^8=K_8,   \label{SU3DHN}
\eea
where the fields $K_i,Q_i,S_i$ depend on space-time coordinates $(r,\theta,t)$. 
The ansatz is consistent with the Euler equations obtained from the
Lagrangian ${\cal L}_{gen}$ and leads to a system of fourteen partial 
differential equations for the field variables $K_i,Q_i,S_i$.
Note, that due to the introduced gauge fixing terms 
the equations for $K_i,Q_i,S_i$ do not admit any residual symmetry
and represent well-defined second order hyperbolic differential equations.

One can simplify further the system of equations for $K_i,Q_i,S_i$
applying the following reduction ansatz
 \bea
&& Q_{1,2,5}=-S_{1,2,5}=-K_{1,2,5},  \nn \\
&& Q_{4}= \big (-\frac{1}{2}+\frac{\sqrt 3}{2}\big ) K_4,  \nn \\
&& S_{4}= \big (-\frac{1}{2}-\frac{\sqrt 3}{2}\big ) K_4,  \nn \\
&& K_{3,8}= -\frac{\sqrt 3}{2} K_4, \label{reduction1}
 \eea
with setting the parameters $\alpha_2=\alpha_5=\alpha_7$.
Without loss of generality we choose $\alpha_a=1$.
Direct substitution of the reduction ansatz into the equations for $K_i,Q_i,S_i$ 
leads to only five linearly independent differential equations
which contain four second order hyperbolic equations for the fields $K_{1,2,4,5}(r,\theta,t)$
and one quadratic constraint with first order partial derivatives
\bea
&r^2 \pro_t^2 K_1-r^2 \pro_r^2 K_1-\pro_\theta^2 K_1+2r(\pro_t K_5-\pro_r K_1)\nn \\
&+\cot\theta(\pro_r K_2-\pro_\theta K_1)+\dfrac{9}{2}\csc^2\theta K_4^2 K_1=0, \label{eq1}
\eea
\bea
&r^2 \pro_t^2 K_2-r^2 \pro_r^2 K_2-\pro_\theta^2 K_2+r^2 \cot\theta(\pro_t K_5-\pro_r K_1)\nn \\
&-\cot\theta \pro_\theta K_2+\dfrac{9}{2}\csc^2\theta K_4^2 K_2=0,\label{eq2}
\eea
\bea
&r^2 \pro_t^2 K_4-r^2 \pro_r^2 K_4-\pro_\theta^2 K_4+\cot\theta\pro_\theta K_4 \nn \\
& +3r^2(K_1^2-K_5^2)K_4+3K_2^2K_4=0,\label{eq3}
\eea
\bea
&r^2 \pro_t^2 K_5-r^2 \pro_r^2 K_5-\pro_\theta^2 K_5+2r(\pro_t K_1-\pro_r  K_5)\nn \\
&+\cot\theta (\pro_t K_2-\pro_\theta K_5)+ \dfrac{9}{2}\csc^2\theta K_4^2 K_5=0, \label{eq4}
\eea
\bea
&2 r^2 (K_5\pro_t K_4-K_1 \pro_r K_4 )+ K_2 (\cot \theta K_4-2 \pro_\theta K_4) \nn \\
&+ K_4 (-\pro_\theta K_2 +r^2 (\pro_t K_5-\pro_r K_1))=0.  \label{constr1}
\eea
The system of equations (\ref{eq1}-\ref{constr1}) is not suitable for numeric solving
due to the presence of the constraint with derivatives and lack of explicit functions defining the boundary
conditions on two-dimensional surfaces closing the three-dimensional numeric domain. 
 To overcome this problem we employ the same method
as in the case of $SU(2)$ QCD using the Fourier series decomposition
for the fields $K_i,Q_i,S_i$ as in  (\ref{seriesdec}) (the Abelian potential $K_{8}$ has
a decomposition similar to one for $K_3$). After integration over the time period in the classical action
one can derive the Euler equations for the field modes $\tK_i^{(n)},\tQ_i^{(n)},\tS_i^{(n)}$
depending on two coordinates $(r,\theta)$. We apply a reduction ansatz to
the obtained Euler equations for $\tK_i^{(n)},\tQ_i^{(n)},\tS_i^{(n)}$
 \bea 
&& \tQ_{1,2,5}^{(n)}=-\tS_{1,2,5}^{(n)}=-\tK_{1,2,5}^{(n)},  \nn \\
&& \tQ_{4}^{(n)}= \big (-\frac{1}{2}+\frac{\sqrt 3}{2}\big ) \tK_4^{(n)},  \nn \\
&& \tS_{4}^{(n)}= \big (-\frac{1}{2}-\frac{\sqrt 3}{2}\big ) \tK_4^{(n)}, \nn \\
&&\tK_{3,8}^{(n)}= -\frac{\sqrt 3}{2} \tK_4^{(n)}, \label{reduction2}
 \eea
 where ($n=1,3,5, ...$), and we impose a condition that all even 
 modes  $\tK_i^{(2k)},\tQ_i^{(2k)},\tS_i^{(2k)}$ vanish. Such a condition
resolves the constraint (\ref{constr1}) and reduces 
 the space of general solutions to the subspace of solutions with a definite parity 
 under the reflection $\theta \rightarrow \pi-\theta$.
It is remarkable that the ansatz (\ref{reduction2}) reduces the total number of
equations to four independent second order hyperbolic equations 
in each order of the Fourier series decomposition  (\ref{seriesdec})
without any additional constraints. 

   Note that the Abelian potentials $\tK_{3,8}$ are 
 equal to each other as it takes place in the case of Abelian Weyl symmetric homogeneous 
 magnetic fields providing an absolute minimum of the quantum effective potential \cite{flyvb,pakMPLA06}. 

We demonstrate existence of another type of monopole pair solution
which is different from the $SU(2)$ stationary monopole pair considered above. 
To find such a solution we apply asymptotic conditions (\ref{asym1}) 
for the functions $\tilde K_{i}^{(n)}$ containing 
even angle modes $a_{2,4}^{(n)}(\theta)$
and odd angle modes $a_{1,5}^{(n)}(\theta)$. 
 Imposing vanishing Dirichlet boundary conditions at the origin
 and along the boundaries $(\theta=0,\pi)$, we solve
the equations for $\tK_{i}^{(n)}$ in the fifth order
approximation. The obtained solution profile functions
$\tK_i^{(1)}$ and energy density plots are presented in Fig. 2. 
\begin{figure}[h!]
\centering
\subfigure[~]{\includegraphics[width=42mm,height=30mm]{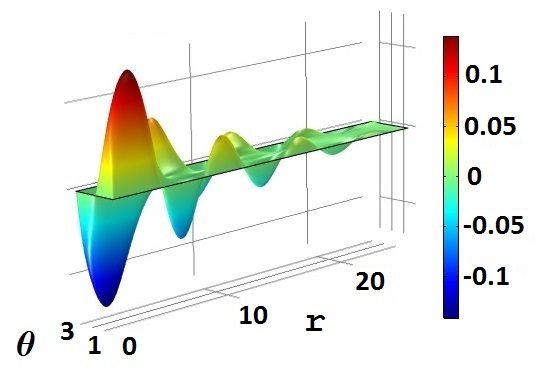}}
\subfigure[~]{\includegraphics[width=42mm,height=30mm]{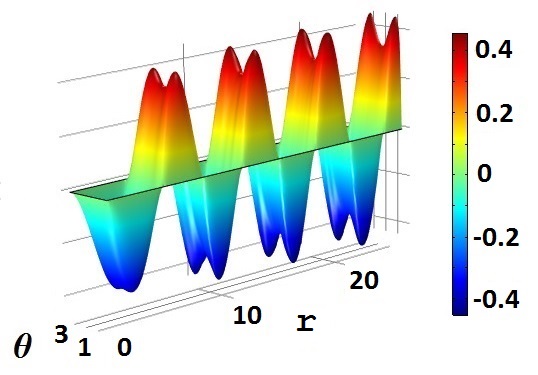}}
\subfigure[~]{\includegraphics[width=42mm,height=30mm]{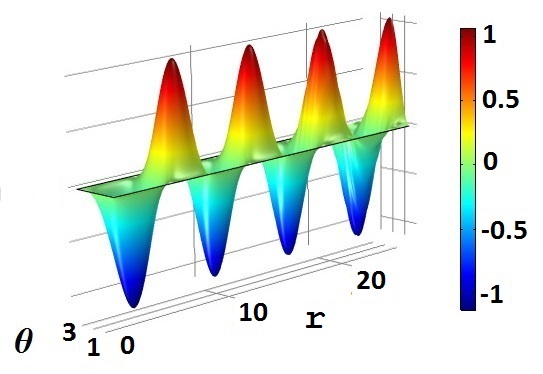}}
\subfigure[~]{\includegraphics[width=42mm,height=30mm]{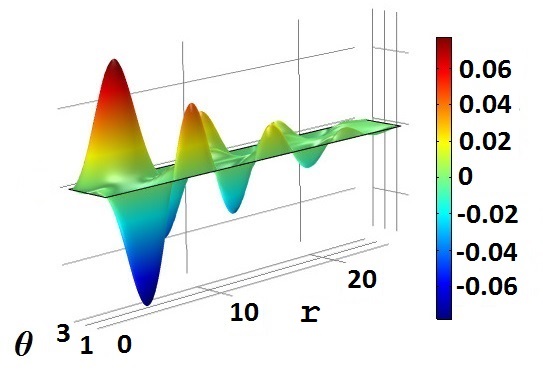}}
\subfigure[~]{\includegraphics[width=42mm,height=36mm]{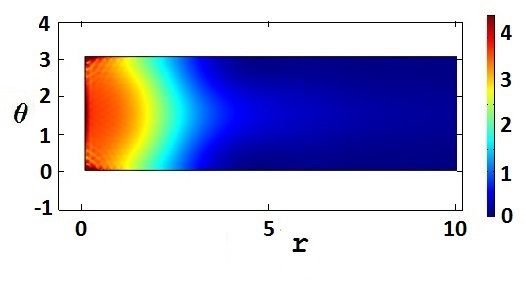}}
\hfill
\subfigure[~]{\includegraphics[width=42mm,height=36mm]{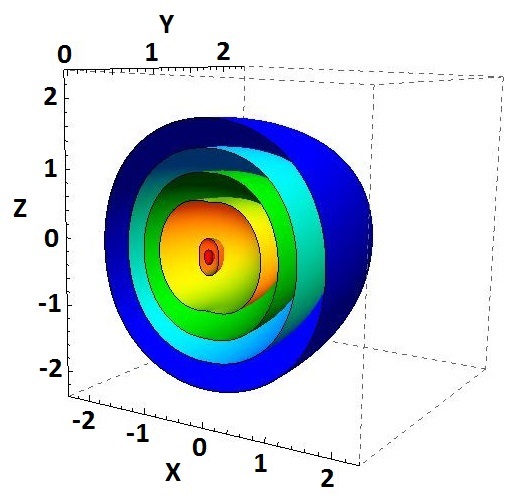}}
\caption[fig2]{Profile functions for the monopole pair solution in the leading order: (a) $\tK_{1}^{(1)}$;
(b)  $\tK_{2}^{(1)}$; (c)  $\tK_{4}^{(1)}$; (d)  $\tK_{5}^{(1)}$; (e) the energy density plot; 
(f) the energy density contour plot in Cartesian coordinates ($g=1,M=1$).}\label{Fig2}
\end{figure}

Consider the magnetic flux structure of the obtained solution
when the observation time is much larger than the period
of time oscillations.
The time averaged radial components of the 
Abelian field strength in the leading order include only
non-linear terms
\bea
\langle F_{\theta\varphi}^{3}\rangle_t=
-\dfrac{3}{4} \tK_2^{(1)} \tK_4^{(1)},&&
\langle F_{\theta\varphi}^{8}\rangle_t =
+\dfrac{3}{4} \tK_2^{(1)} \tK_4^{(1)},
\eea
where the Abelian potential $\tK_4^{(1)}$ contains the angle dependent factor 
$\sin^2 \theta$, and $\tK_2^{(1)}$ is an even function with respect to the reflection symmetry
$\theta\rightarrow \pi-\theta$.
The magnetic fields $\tF_{\theta\varphi}^{3,8(1)}$ 
create opposite magnetic fluxes through a sphere of radius $R$ with a center at the
origin  which correspond to non-topological antimonopole and monopole located 
at one point. 
Careful numeric analysis shows that stationary solutions corresponding to
a lowest energy in a chosen finite numeric domain are classified 
by only two parameters, the conformal parameter $M$ and an amplitude $c_0$ of the 
oscillating Abelian potential $\tK_4^{(1)}$.
A detailed structure of the numeric solution up to the fifth order 
series decomposition is given in \cite{SM}. 

\section{Microscopic quantum stability of the stationary solutions}

The Savvidy QCD vacuum based on the classical homogeneous color magnetic field
is unstable due to the presence of an imaginary part of the effective action \cite{savv, N-O}.
Usually one expects that introducing time dependent color fields as vacuum makes worse the 
vacuum stability since the color electric field leads to an imaginary part of the effective action 
as well \cite{schan82}. Surprisingly, it has been found that non-linear plane wave solutions 
make the problem of vacuum stability more soft, in a sense, that an equation for 
the unstable modes is very similar to the equation for an electron in the periodic potential \cite{ijmpa2017}. 
This gives a hint that one can find a proper stationary periodic
wave type solution which provides a stable vacuum. Indeed, recently it has been proved that
a stationary spherically symmetric generalized Wu-Yang monopole solution leads to a stable vacuum
\cite{prd2017}. 
We will prove the quantum stability of the stationary
axially-symmetric solutions under small quantum gluon fluctuations
in the case of $SU(2)$ and $SU(3)$ QCD. 

To verify the stability of the solutions it is suitable to apply the quantum 
effective action formalism. We consider one-loop quantum effective action
expressed in terms of functional operators \cite{chopakprd02}
\bea
S^{\rm 1 loop}_{\rm eff}&=&-\dfrac{1}{2} {\rm Tr} \ln [K_{\mu\nu}^{ab}]+
{\rm Tr}\ln [M_{\rm FP}^{ab}], \nn \\
K_{\mu\nu}^{ab}&=&-
\delta^{ab} \delta_{\mu\nu} \pro^2_t
            -\delta_{\mu\nu}({\Da}_\rho {\Da}^\rho)^{ab} -2 f^{acb}{\mathcal
	    F}_{\mu\nu}^c, \quad \label{Koper}\\
M_{\rm FP}^{ab}&=&-({\Da}_\rho{\Da}^\rho)^{ab},   \nn 
\eea
where the Wick rotation $t \rightarrow -i t$
has been performed to provide the causal structure of the operators,
and ${\Da}_\mu,{\mathcal F}_{\mu\nu}^c $ 
are defined with a classical background field $\Ba_\mu^a$. 
The operators $K_{\mu\nu}^{ab}$ and $M^{ab}_{\rm FP}$ correspond to
gluon and Faddeev-Popov ghosts. 
Note that the expression (\ref{Koper}) is valid for arbitrary background field
and does not depend on a chosen gauge for the background field
due to use of a gauge covariant background formalism \cite{abbott}.
Effective action describes the vacuum-vacuum amplitude, and the presence of 
an imaginary part of the action implies vacuum instability.
Therefore, if the operator $K_{\mu\nu}^{ab}$ is not positively defined then
an unstable mode will appear as an eigenfunction corresponding to a negative eigenvalue
of the following  ``Schr\"{o}dinger'' equation 
\bea
K_{\mu}^{ab} \Psi_\nu^b=\lambda \Psi_\mu^a, \label{schr3}
\eea
where the ``wave functions'' $\Psi_\mu^a(t, r,\theta,\varphi)$ describe
gluon fluctuations. Note that the ghost operator $M^{ab}_{\rm FP}$ is positively defined and does not
produce instability \cite{N-O}. The potential in the operator $K_{\mu}^{ab}$ does not
depend on the asimuthal angle. Due to this one can separate a corresponding angle
dependent part from the function $\Psi_\mu^a$ and solve the eigenvalue equation
in a three-dimensional domain $(t,r,\theta)$.
Substituting interpolation functions for the stationary
magnetic solutions in the leading order, one can 
solve the eigenvalue equation (\ref{schr3}). 
In the case of  $SU(2)$ stationary solution
a full eigenvalue spectrum is divided into
four sub-spectra corresponding to four decoupled systems 
of equations: 
(I) $\Psi_4^2$, (II) $\Psi_4^1,\Psi_4^3$, (III) $\Psi_ 1^2,\Psi_2^2, \Psi_3^1,\Psi_3^3$,
(IV) $\Psi_1^1,\Psi_1^3,\Psi_2^1,\Psi_2^3,\Psi_3^2$.
The lowest eigenvalue is positive, and it is reached by a solution satisfying the system
of equations (II), the corresponding eigenfunctions are plotted in Fig. 3  ($g=1,M=1$).
 \begin{figure}[h!]
\centering
\subfigure[~]{\includegraphics[width=42mm,height=32mm]{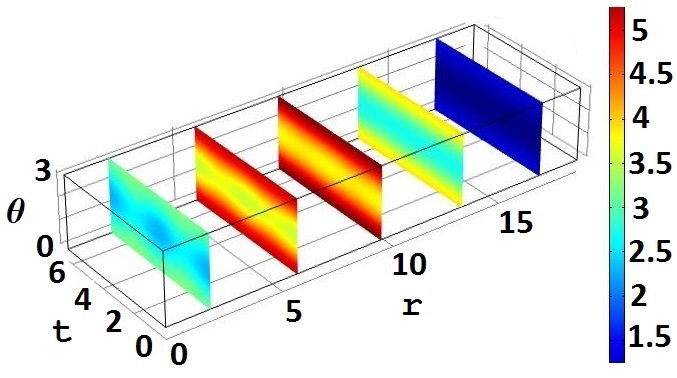}}
\subfigure[~]{\includegraphics[width=42mm,height=32mm]{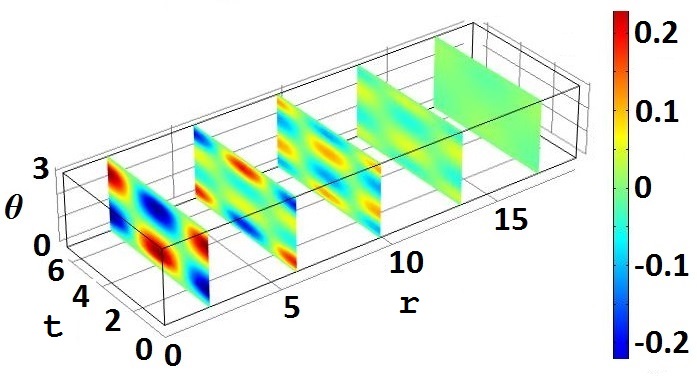}}
\caption[fig3]{Non-vanishing eigenfunctions corresponding to the lowest
eigenvalue $\lambda=0.05294$: (a) $\Psi_4^1$; 
 (b) $\Psi_4^3$.
}\label{Fig3}
\end{figure}
Other systems of equations, (I,III,IV),  have a similar structure of the eigenvalue spectrum. 

In the case of $SU(3)$ stationary monopole pair solution
the equations in (\ref{schr3}) are not factorized, 
and one has to solve a full set of thirty two differential equations. 
Numerical results of solving the  ``Schr\"{o}dinger'' eigenvalue equation 
with the stationary monopole-antimonopole background field
show that the eigenvalue spectrum is positively defined. 
The obtained dependence of the lowest eigenvalue on the size of the chosen
numeric domain for large values of $L$ confirms the positiveness of the 
eigenvalue spectrum, Fig. 4. 
This proves the quantum stability of stationary monopole-antimonopole pair solutions in QCD.
 \begin{figure}[h!]
\centering
\includegraphics[width=80mm,height=50mm]{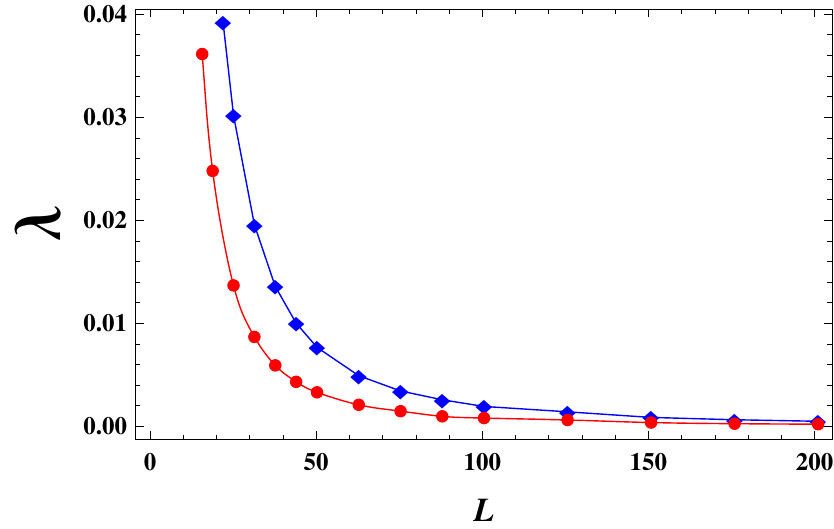}
\caption[fig4]{Dependence of the ground state eigenvalue $\lambda(L)$ 
on the size $L$ of the numeric domain
in the cases of $SU(3)$ (red curve) and $SU(2)$ (blue curve) stationary monopole-antimonopole
pair solutions  ($g=1,M=1$).}\label{Fig4}
\end{figure}
The quantum stability of $SU(2)$ and $SU(3)$ stationary background fields has been
checked for solutions with amplitude values
of the Abelian potential $K_{4}(r,\theta,t)$ in the interval $(0\leq c_0 \leq 2)$
and with conformal parameter values ($0\leq M \leq 2$). For large values 
of $M,c_0$ unstable modes appears which destabilize the vacuum.

\section{Weyl symmetry and microscopic structure of the QCD vacuum}

Let us consider symmetry properties of essentially $SU(3)$ stationary solutions.
The classical Yang-Mills Lagrangian can be 
rewritten in terms of Weyl symmetric fields \cite{pakMPLA06}
\bea
{\cal L}_0&=&\sum_{p=1,2,3}\Big \{  -\dfrac{1}{6} ({\cal G}|_{\mu\nu}^p)^2-\dfrac{1}{2}|D_\mu^p W_\nu^p-D_\nu^p W_\mu^p|^2  \nn \\
&&-ig {\cal G}_{\mu\nu}^p W_\mu^{*p} W_\nu^p\Big \}-{\cal L}_{int}^{(4)}[W] , \label{WeylLagr}
\eea
where ${\cal G}_{\mu\nu}^p$ are Abelian field strengths containing
the gauge potentials $A_{\mu}^{3,8}$,
the complex fields $W_\mu^p$ represent off-diagonal gluons, and 
the index $p$ counts the Weyl symmetric gauge potentials. 
Note that the Lagrangian is not Weyl symmetric under permutation of $I,U,V$ subgroups
$SU(2)$ since the quartic interaction term ${\cal L}_{int}^{(4)}[W]$ is not factorized into a 
sum of separate parts corresponding to $I,U,V$ sectors. 
It is remarkable that the Lagrangian on the space of essentially $SU(3)$ stationary
solutions possesses a high symmetric structure. First of all, substituting the reduction ansatz
(\ref{reduction1}) for general functions $K_i,Q_i,S_i$ into the Lagrangian, one can verify that 
${\cal L}_{int}^{(4)}$ obtains an explicit Weyl symmetric form
\bea
{\cal L}_{int}^{(4)}[W]=\dfrac{9}{8}\sum_{p=1,2,3} \Big ( (W^{*p}_\mu W^p_\mu)^2-(W^*_\mu)^2 (W_\nu)^2
\Big ) 
\eea
Secondly, each $I,U,V$ sector in the Lagrangian (\ref{WeylLagr}) contains
cubic interaction terms corresponding to the anomaly magnetic moment interaction
which is precisely the source of the Nielsen-Olesen vacuum instability \cite{N-O}.
 It is surprising, within the framework of the ansatz (\ref{reduction1}) 
one has complete mutual cancellation of all cubic interaction terms. This implies that 
on the space of Weyl symmetric fields the classical action 
describes a generalized $\lambda \phi^4$ theory.

A simple consideration shows that our approach to QCD vacuum problem 
based on stationary Weyl symmetric monopole pair solutions
opens a new perspective towards construction of a microscopic 
theory of vacuum and vacuum phase transitions.
First of all, note that a system of separated stationary generalized 
Wu-Yang monopoles and antimonopoles can not be stable due mutual attraction
between the monopole and antimonopole. In addition, despite on the quantum stability
of a sinlge stationary spherically symmetric monopole, the solution is rather classically 
unstable with respect to small axially-symmetric field deformations \cite{ijmpa2017}.
This implies that axially-symmetric solutions are more preferable as candidates for
the vacuum. Another important feature of the monopole pair solution is that
it represents a non-trivial essentially non-Abelian field configuration
which describes a pair of monopole and antimonopole located at one point. 
This implies that monopole and antimonopole, as well as two monopole-antimonopole pairs with opposite
color orientations, can merge into a stable state with a finite energy density
in the limit of zero distance between the monopole and antimonopole.
In other words, the existence of a stable solution for a pair of monopole and antimonopole
located at one point prevents from annihilation and disappearance of the monopoles.
This is contrary to the case of Dirac and Wu-Yang pair of monopole
and antimonopole which annihilate when they meet each other.

We expect that the QCD vacuum is formed due to condensation of monopole-antimonopole
pairs, and the microscopic vacuum structure is characterized by few parameters:
the conformal parameter $M$, the amplitude of oscillations $c_0$
of the Abelian gauge potential $K_4$ in the asymptotic region,  and the 
concentration of monopole pairs at zero temperature.
Numeric analysis shows that with increasing temperature the internal energy of each monopole pair
increases and at some critical values of the parameters ($M>2,c_0>2$) the 
monopole-antimonopole pair becomes unstable. Note that
in the confinement phase  the vacuum averaging value of the gluon field
operator $\langle0|A_\mu^a|0\rangle$ vanishes since the size of the hadron
is much larger than the characteristic length $\lambda_M=2 \pi/M$
of the vacuum monopole field oscillations. To describe dynamics of the vacuum structure 
at finite temperature one should apply the Euclidean functional integral formalism 
with time integration in the finite interval $(0\leq t \leq \beta=1/kT)$.
It is clear that at high enough temperature the upper integration limit $\beta$ will be less
than $\lambda_M$. This will lead to a non-vanishing vacuum averaging value
of the gluon field operator, $\langle0|A_\mu^a|0\rangle$, and
transition to the deconfinement phase with spontaneous symmetry breaking
where the gluon can be observed as a color object.\\


\section{Conclusion}

In conclusion, we have proposed a new class of regular axially-symmetric stationary solutions
in a pure $SU(2)$ and $SU(3)$ QCD. The solutions possess interesting features such as an intrinsic
mass parameter, a vanishing classical canonical spin density. Such properties serve as a heuristic 
argument to existence of a stable quantum vacuum condensate in the quantum theory.
After time averaging over the period
the solutions correspond to color magnetic field configurations which have asymptotic behavior 
similar to one of the non-Abelian monopole-antimonopole pair. A careful numeric analysis 
confirms stability of the stationary solutions under small gluon fluctuations within the framework
of one-loop effective action formalism. As it is known, in QCD the quantum dynamics
leads to generation of the mass gap, or the so-called vacuum gluon condensate parameter.
So that, the mass scale parameter $M$ of the classical stationary solutions is related to
the finite mass gap parameter and characterizes the microscopic scale of the vacuum structure.
The presence of such a parameter allows to describe phase transitions in QCD. The most 
important step in construction of the full microscopic theory of the QCD vacuum is
to study condensation of monopole-antimonopole pairs. This will be considered in a separate paper.


 \acknowledgments
 
 One of authors (DGP) thanks Prof. C.M. Bai
for warm hospitality during his staying in Chern Institute of Mathematics
and Dr. Ed. Tsoy for useful discussions of numeric aspects.
The work is supported by:
(YK) Rare Isotope Science Project of Inst. for Basic Sci. funded by Ministry of Science, 
ICT and Future Planning, and National Reserach Foundation of Korea, grant NRF-2013M7A1A1075764; 
(BHL) NRF-2014R1A2A1A01002306 and NRF-2017R1D1A1B03028310; 
(CP) Korea Ministry of Education, Science and Technology, Pohang city, and NRF-2016R1D1A1B03932371;
(DGP) Korean Federation of Science and Technology, Brain Pool Program, and grant OT-$\Phi$2-10.


\end{document}